\documentclass[twocolumn]{aa}
\usepackage{graphicx}
\usepackage{txfonts}
\def\kms{\relax \ifmmode {\,\rm km\,s}^{-1}\else \,km\,s$^{-1}$\fi}
\def\ks{\relax \ifmmode  K_{\rm s}\else $K_{\rm s}$\fi}
\def\ha{\relax \ifmmode {\rm H}\alpha\else H$\alpha$\fi}
\def\hb{\relax \ifmmode {\rm H}\beta\else H$\beta$\fi}
\def\hi{\relax \ifmmode {\rm H\,{\sc i}}\else H\,{\sc i}\fi}
\def\hii{\relax \ifmmode {\rm H\,{\sc ii}}\else H\,{\sc ii}\fi}
\def\h2{\relax \ifmmode {\rm H}_2\else H$_2$\fi}
\def\lha{\relax \ifmmode L_{{\rm H}\alpha}\else $L_{{\rm H}\alpha}$\fi}
\def\shi{\relax \ifmmode \sigma_{{\rm HI}}\else $\sigma_{\rm HI}$\fi}
\def\sh2{\relax \ifmmode \sigma_{{\rm H}_2}\else $\sigma_{{\rm H}_2}$\fi}
\def\degr{\hbox{$^\circ$}}
\def\arcmin{\hbox{$^\prime$}}
\def\arcsec{\hbox{$^{\prime\prime}$}}

\def\fdg{\hbox{$.\!\!^\circ$}}
\def\fs{\hbox{$.\!\!^{\rm s}$}}
\def\farcm{\hbox{$.\mkern-4mu^\prime$}}
\def\farcs{\hbox{$.\!\!^{\prime\prime}$}}
\def\degd#1.#2{ #1\fdg#2 }                 
\def\mind#1.#2{ #1\farcm#2 }               
\def\secd#1.#2{ #1\farcs#2 }               
\def\hhh{\ifmmode {\rm ^h}              
         \else {${\rm ^h}$}
         \fi}
\def\sss{\ifmmode {\rm ^s}              
         \else {${\rm ^s}$}
         \fi}
\def\hms#1h#2m#3s{                      
                  \relax
                  \ifmmode #1^{\rm h}\,#2^{\rm m}\,#3^{\rm s}
                  \else \hbox{$#1^{\rm h}\,#2^{\rm m}\,#3^{\rm s}$}
                  \fi
                 }
\def\dms#1d#2m#3s{                      
                  \relax
                  #1\degr\,#2\arcmin\,#3\arcsec 
                 }
\def\hmsd#1h#2m#3.#4s{                  
                      \relax
                      \ifmmode #1^{\rm h}\,#2^{\rm m}\,#3\fs#4
                      \else \hbox{$#1^{\rm h}\,#2^{\rm m}\,#3\fs#4$}
                      \fi
                     }
\def\dmsd#1d#2m#3.#4s{                  
                      \relax
                      #1\degr\,#2\arcmin\,#3\farcs#4
                     }
\def\mag{\relax                          
        \ifmmode ^{\rm m}
        \else $^{\rm m}$
        \fi
       }
\def\magd#1.#2{                          
              \relax
              \ifmmode #1^{\rm m}
                       \hskip-0.55em.\hskip0.22em#2
              \else \hbox{#1$^{\rm m}
                    \hskip-0.55em.\hskip0.22em$#2}
              \fi
             }
\input psfig.sty
\begin{document}
\title{The nuclear ring in the unbarred galaxy NGC 278: result of a minor 
merger?}  
\author{J.  H.  Knapen\inst{1} \and L.~F.~Whyte\inst{2} \and
W.~J.~G. de Blok\inst{3} \and J.~M. van der Hulst\inst{4} }

\titlerunning{The nuclear ring in the unbarred galaxy NGC 278}

\offprints{J. H. Knapen}  

\institute{Centre for Astrophysics Research, University of
Hertfordshire, Hatfield, Herts AL10 9AB, U.K.\\
\email{j.knapen@star.herts.ac.uk} 
\and University of Nottingham,
School of Physics and Astronomy, University Park, Nottingham, NG7 2RD,
U.K.
\and Department of Physics and Astronomy, Cardiff University, 5 The Parade, 
Cardiff CF24 3YB, U.K.
\and Kapteyn Astronomical Institute, University of Groningen, Postbus 800, 
NL-9700 AV Groningen, The Netherlands
}

\date{Received ; accepted May 4, 2004}

\abstract{

We present fully sampled high angular resolution two-dimensional
kinematics in the \ha\ spectral line, optical and near-infrared
imaging, as well as 21~cm atomic hydrogen data of the spiral galaxy
NGC~278.  This is a small non-barred galaxy, which has a bright star
forming inner region of about 2~kpc in diameter, reminiscent of nuclear
rings seen mainly in barred galaxies.  The \ha\ kinematics show a
disturbed velocity field, which may be partly the result of spiral
density wave streaming motions.  The 21~cm data trace the atomic
hydrogen well outside the optical disk.  The \hi\ is not abundant but
clearly shows disturbed morphology and kinematics.  We postulate that
the current structure of NGC~278 is a result of a recent interaction
with a small gas-rich galaxy, which is now dispersed into the outer
disk of NGC~278.  Non-axisymmetries set up in the disk by this
minor merger may well be the cause of the intense star formation in the
inner region, which can be interpreted as a rare example of a nuclear
ring in a non-barred galaxy. Rather than being induced by a bar, this
nuclear ring would then be the direct result of an interaction event
in the recent history of the galaxy.


}

\maketitle

%

\section{Introduction}

Rings in disk galaxies are intimately linked to the internal dynamics
and the evolution of their hosts. Rings are common, are more often
than not sites of recent massive star formation (SF), and are linked
observationally and phenomenologically to the presence of bars (see
Buta \& Combes 1996 for a general review on galactic rings).  Nuclear
rings occur in about one fifth of spiral galaxies (Knapen 2004).  They
are found almost exclusively in barred galaxies (e.g., Buta \& Combes
1996; Knapen 2004), and can be directly linked to inner Lindblad
resonances (ILRs), between the epicyclic oscillations of the stars and
the rotation of the (bar) pattern (Knapen et al. 1995a; Heller \&
Shlosman 1996; Shlosman 1999). Because of the much enhanced massive SF
occurring in nuclear rings, they are prime tracers of SF processes in
starburst regions, as well as of the dynamics of galaxies close to the
nucleus. Although nuclear rings are intimately connected to bar
dynamics, in rare cases a nuclear ring occurs in a galaxy where no bar
can be identified. In this paper, we study one such case, that of
NGC~278, in detail.

We describe high resolution \ha\ Fabry-P\'erot (FP) data, in
combination with optical and near-infrared (NIR) imaging and with \hi\
21~cm synthesis data, of the rather face-on SAB(rs)b (de Vaucouleurs
et al.  1991, hereafter RC3) spiral galaxy NGC~278 (UGC~528).  NGC~278
has several knotty spiral arms, first described by Pease (1917).
Pease also noticed the sudden drop in intensity at about 50~arcsec in
diameter, which is just outside the region where the much enhanced SF
is organised into a nuclear ring (e.g., Pogge 1989; Sect.~4.5) of
about 40~arcsec, or some 2.2~kpc, in diameter.  The galaxy is small
($D_{25}=2.1$~arcmin, RC3; or a diameter of 7.2~kpc, Tully 1988) but
has an \hi\ disk which extends significantly further out, as we will
show below.  At a distance of 11.8~Mpc (Tully 1988), 1~arcsec
corresponds to 57~pc.

NGC 278 is part of the local supercluster of galaxies centred around
the Virgo cluster (Vallee 1993).  Arp \& Sulentic (1985), in their
analysis of groups of galaxies, listed NGC 278 as having a companion
dwarf galaxy UGC 672. In their study of interacting galaxies,
Bushouse, Werner \& Lamb (1988), however, considered NGC 278 to be an
isolated galaxy, which is in agreement with Karachentseva et
al. (1996) who found UGC~672 to be an isolated dwarf galaxy.  Although
the difference in systemic velocity between the two galaxies is only
68\kms\ (RC3), combining their positions on the sky with the scale as
given in the previous paragraph, we find that NGC~278 and UGC~672 are
some 600~kpc distant, too much to consider them interacting,
especially when considering the faint absolute magnitude of UGC~672,
of about $-12.4$ (estimated using parameters from the RC3, in
comparison, $M_B=-18.8$ for NGC~278).  We will come back to
interactions later in this paper.

Pogge (1989) reported a round, 11~arcsec (450~pc using our distance
assumption) diameter, nuclear ring ``embedded in a bright,
starbursting disk'' from \ha\ imaging. Deconvolved $F555W$ (roughly
$V$-band) images of NGC~278 taken with the {\it Hubble Space Telescope
(HST)} WFPC revealed evidence for the existence of spiral structure to
within $2-3$~arcsec of the unresolved nucleus, and for a narrow dust
lane approaching within 0.5 arcsec of the very centre (Phillips et al.
1996).  Schmidt, Bica \& Alloin (1990) used a combination of optical
spectroscopy and stellar population synthesis modelling to study the
central region of NGC~278, and suggest that the predominant H{\sc ii}
regions are in late evolutionary stages.  The high luminosity of the
galaxy is almost certainly caused by intense SF, a burst of which is
still operating (Schmidt et al. 1990). More recently, Garrido et
al. (2003) published FP observations of, among other galaxies,
NGC~278, obtained as part of their GHASP survey. Their data, at a
spatial resolution of 2~arcsec or almost twice as large as our data,
show roughly the same morphological and velocity structure as
presented in the current paper. Garrido et al. (2003) claim that
NGC~278 shows a barred spiral structure, although it is not clear on
the basis of which source they may have reached this conclusion, and
also that the isovelocity contours confirm the presence of a bar. We
will show below, on the basis not only of FP \ha\ data but also of a
large quantity of optical and NIR imaging, as well as the WHISP \hi\
data to which Garrido et al. briefly refer in their paper, that
NGC~278 does not show any evidence for the presence of a bar, and that
the behaviour of its gaseous velocity field, as well as the presence
of the nuclear ring of SF, is a likely result of a past interaction.

After describing the FP and \hi\ data and our optical, NIR and {\it
HST} imaging (Sect.~2), we detail the morphological and kinematic
results in Sect.~3.  We discuss the spiral structure, nuclear ring,
and the overall morphology, as well as the plausible recent history of
the galaxy, in Sect.~4, and summarise our main results in Sect.~5.

\section{Observations and data reduction}

\subsection{Fabry-P\'erot data}

The \ha\ FP maps were obtained on the night of 1998 September 2 with
the TAURUS-II instrument, equipped with a TEK CCD camera, on the 4.2m
William Herschel Telescope (WHT) on La Palma. The data reduction
procedures are summarised below, but are very similar to those used by
Knapen et al.  (2000) and Laine et al.  (2001), to which we refer for
further details.  For the data set of NGC~278, the CCD was windowed to
400 pixels $\times400$ pixels of 0.28 arcsec pixel$^{-1}$.  The night
was photometric with sub-arcsecond seeing.  We used an appropriately
redshifted narrow band H$\alpha$ filter as an order-sorting filter
($\lambda_{\rm c}$=6577\AA, $\Delta$$\lambda$=15\AA, using the
galaxy's systemic velocity $v_{\rm sys}$=640\kms).  Wavelength and
phase calibration were performed using observations of a calibration
lamp before and after each science exposure. The final calibrated data
cube has 400 pixels $\times$ 400 pixels $\times$ 55 ``planes'' in
wavelength, the latter separated by 0.405 \AA, or 18.5\kms.  The
spatial resolution of the resulting data set is $\sim$1.1~arcsec.  The
data set was placed on an astrometrically correct grid by comparing
the positions of point-like features in the \ha\ FP data with those in
the broad-band images (see below), which in turn could be referenced
to astrometry from a digitised sky survey image using fits to
positions of foreground stars.

\begin{figure*}
\caption{Channel maps of the H$\alpha$ emission from NGC~278 at
2~arcsec resolution. Velocity in \kms\ of each channel is indicated in
the upper left corner. Contour and grey-scale levels are 3 to 15
$\sigma$ in steps of 3 $\sigma$, with an additional contour at
$-3\sigma$.  The kinematic centre of the galaxy is marked with a
cross.}
\label{chan}
\end{figure*}

We determined which channels of the data set were free of \ha\ line
emission after smoothing it to a resolution of
4~arcsec\,$\times$\,4~arcsec.  Subsequently, we subtracted the continuum
emission after fitting the continuum to the 15 line-free channels on
either extreme of the data cube. The \ha\ emission in the slightly
smoothed data set (to 2 arcsec spatial resolution), as a function of
increasing wavelength, or velocity, is shown as a set of channel maps
in Fig.~\ref{chan}. In order to increase the signal-to-noise ratio
across the region of the galaxy where \ha\ emission is detected, we
smoothed the original data with Gaussians to produce a number of
similar data sets at spatial resolutions of 1.2, 1.5, 2, 3 and 4
arcsec.

\begin{figure*}
\psfig{figure=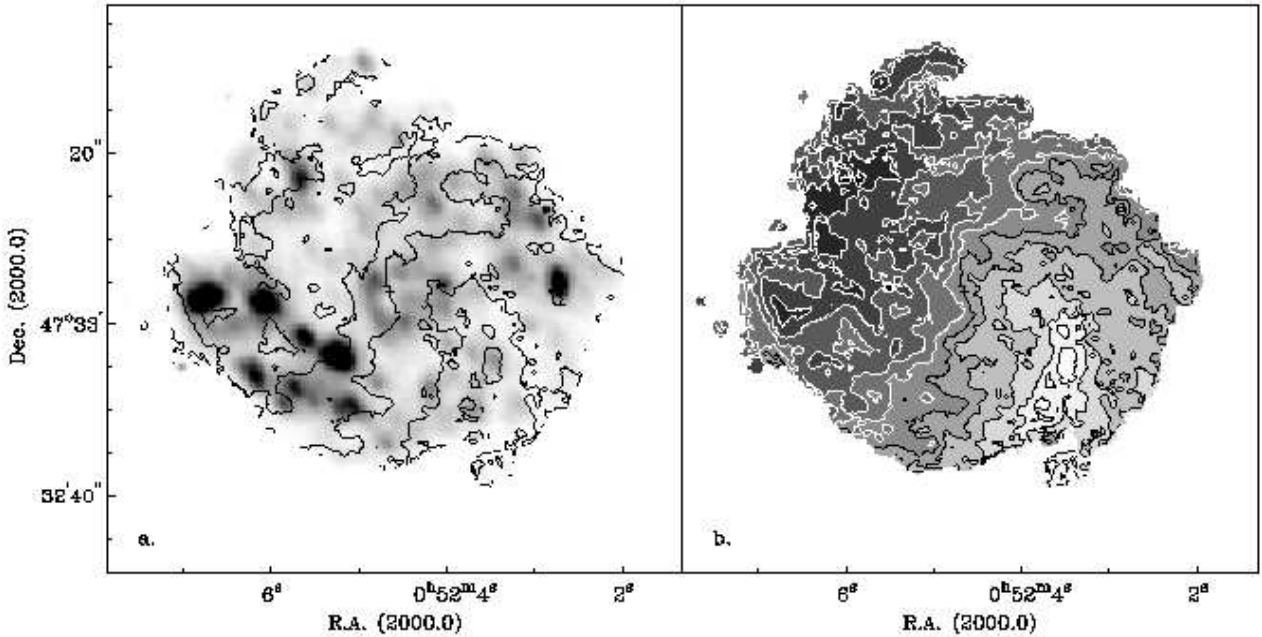,width=17cm,angle=0}
\caption{Moment maps of the FP data set of NGC~278 at 1.2~arcsec
resolution: H$\alpha$ total intensity map (left) and velocity field
(right).  Grey-scale and contour levels in the right panel are from
600 to 690\kms\ in steps of 10\kms, with the lower velocity contours
in black, and the first white contour at 650\kms\ (darker grey shades
thus indicate higher velocities).  In the left panel, only every
second velocity contour is shown. The kinematic centre of the galaxy
is marked with a cross.}
\label{moms}
\end{figure*}

Following the procedures described in detail by Knapen (1997) and
Knapen et al. (2000), we produced moment maps of the data cubes at
different spatial resolution. Moment zero (\ha\ total intensity) and
one (velocity field) maps at a spatial resolution of 1.2~arcsec are
shown in Fig.~\ref{moms}.

\subsection{NIR imaging}

\begin{figure*}
\caption{NIR $J, H$, and $K'$-band images, with contour levels set at
steps of 0.5~mag, and (bottom right) grey-scale representation of the
$J-K'$ colour index image, with redder regions shown as darker shades.}
\label{nir}
\end{figure*}

We obtained $J, H$ and $K'$ images of NGC~278 on the night of 1995
November 5 using the 3.6~m Canada-France-Hawaii Telescope (CFHT) and
the Montreal NIR camera (MONICA; Nadeau et al. 1994), equipped with a
256 x 256 pixel HgCdTe array detector with a projected pixel size of
0.248~arcsec.  Full details of the image reduction can be found in
P\'erez-Ram\'\i rez et al. (2000). The resolution of the NIR images is
slightly better than 1 arcsec. We show our NIR imaging results in
Fig.~\ref{nir} in the form of grey-scale and contour plots of the $J,
H$ and $K'$ images, and a $J-K'$ colour index image.  The NIR images
have not been photometrically calibrated, but we will assume an
average colour of $J-K=0.85$, as obtained from Frogel et al. (1978)
using an absolute magnitude of the galaxy of $M_B=-18.8$ (derived from
the RC3 and the assumed distance).

\subsection{Ground-based optical imaging}

\begin{figure*}
\caption{Top panels: grey-scale and contour plots of the $R$ and
$B$-band images, with in the $R$-band, black contours from 23.5 to
19.5 in steps of 1~mag\,arcsec$^{-2}$ and white contours at 19.0, 18.5
and 18.0 mag\,arcsec$^{-2}$, and in the $B$-band black contours from
24.8 to 20.8 in steps of 1~mag\,arcsec$^{-2}$ and white contours at
20.3, 19.8 and 19.3 mag\,arcsec$^{-2}$. Lower panels: grey-scale plot
of the $B-R$ colour index image (left), with a closeup of the central
kiloparsec (right). Darker shades are redder.  Vertical lines are due
to a bright star located about 2.5~arcmin to the North of the galaxy.
A cross marks the centre of the galaxy.}
\label{optical}
\end{figure*}

Optical images in the $B$ and $R$ bands were obtained from the Isaac
Newton Group's (ING) data archive.  The images, of 300~s exposure time
each, were obtained on 1995 December 23 with the prime focus camera on
the Isaac Newton Telescope, and were flat-fielded by us using standard
procedures.  The TEK3 CCD gives a pixel scale of 0.59
arcsec\,pixel$^{-1}$, and the seeing as measured on the images was 1.1
arcsec. Photometric calibration was obtained from exposures of the
standard star field PG~0231+051. Grey-scale and contour
representations of the $B$ and $R$ images, and of a $B-R$ colour index
image, are shown in Fig.~\ref{optical}.

These optical images of the galaxy are influenced by stray light, and
CCD charge overflow, from the 8.8~$B$-mag A0 star HD~4950, located
just 2.5 arcmin north of the centre of NGC~278.

\subsection{{\it HST} imaging}

\begin{figure*}
\caption{Real-colour image made from the $F450W$, $F606W$, and $F814W$
{\it HST} images.  Area shown is about 100~arcsec on a side, or about
5.7~kpc. N is up, E to the left. The image is reproduced in black and
white in the printed version of the journal, but is available in
colour in the electronic version.}
\label{hst}
\end{figure*}

From the {\it HST} archive, we obtained images taken through the
$F450W, F606W$, and $F814W$ filters with the WFPC2 camera.  The images
were taken on 2001 August 20, and form part of a snapshot survey
GO~9042, the PI of which is Dr.  Stephen Smartt.  Two images of the
galaxy, of 160~s each, were taken through each filter.  Whereas the
INT images (Sect.~2.3) are deeper and show the faint outer structure
in more detail, the {\it HST} images show the structure in the inner
kpc region in much clearer detail thanks to the high spatial
resolution.  We present the {\it HST} images, combined into a
real-colour view of the galaxy, in Figure~\ref{hst}.

\subsection{HI data}

\begin{figure*}
\psfig{figure=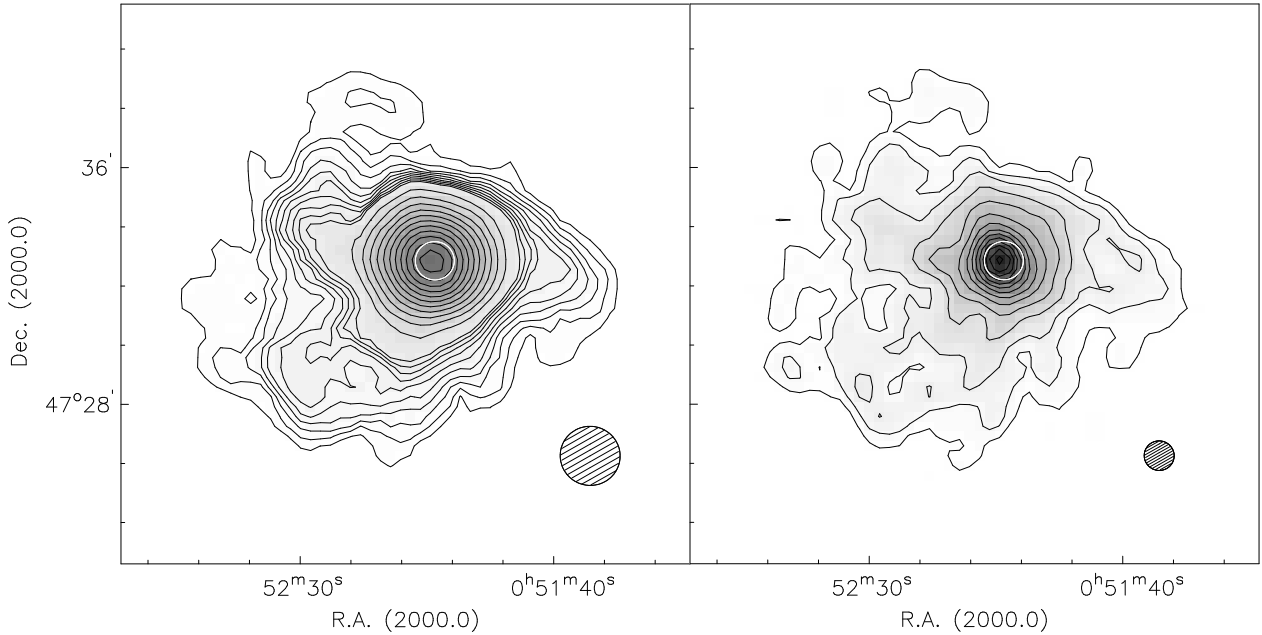,width=17cm,angle=-90}
\caption{Integrated \hi\ surface density distributions at 120~arcsec
(left) and 60~arcsec (right) resolution.  In the left panel the
contour levels are $(1, 2, 3, ...  , 9)\times10^{19}, (1, 1.5, 2, ...,
6.5)\times10^{20}$\,cm$^{-2}$.  The contour levels in the right panel
are $(0.2, 0.5, 1, 2, 3, ..., 11)\times10^{20}$\,cm$^{-2}$. In both
panels, the white contour indicates the extent of the optical
component of NGC~278, as visible on the DSS. FWHM beam size is shown
in the lower right corner of each panel.}
\label{himom0}
\end{figure*}

\begin{figure}
\psfig{figure=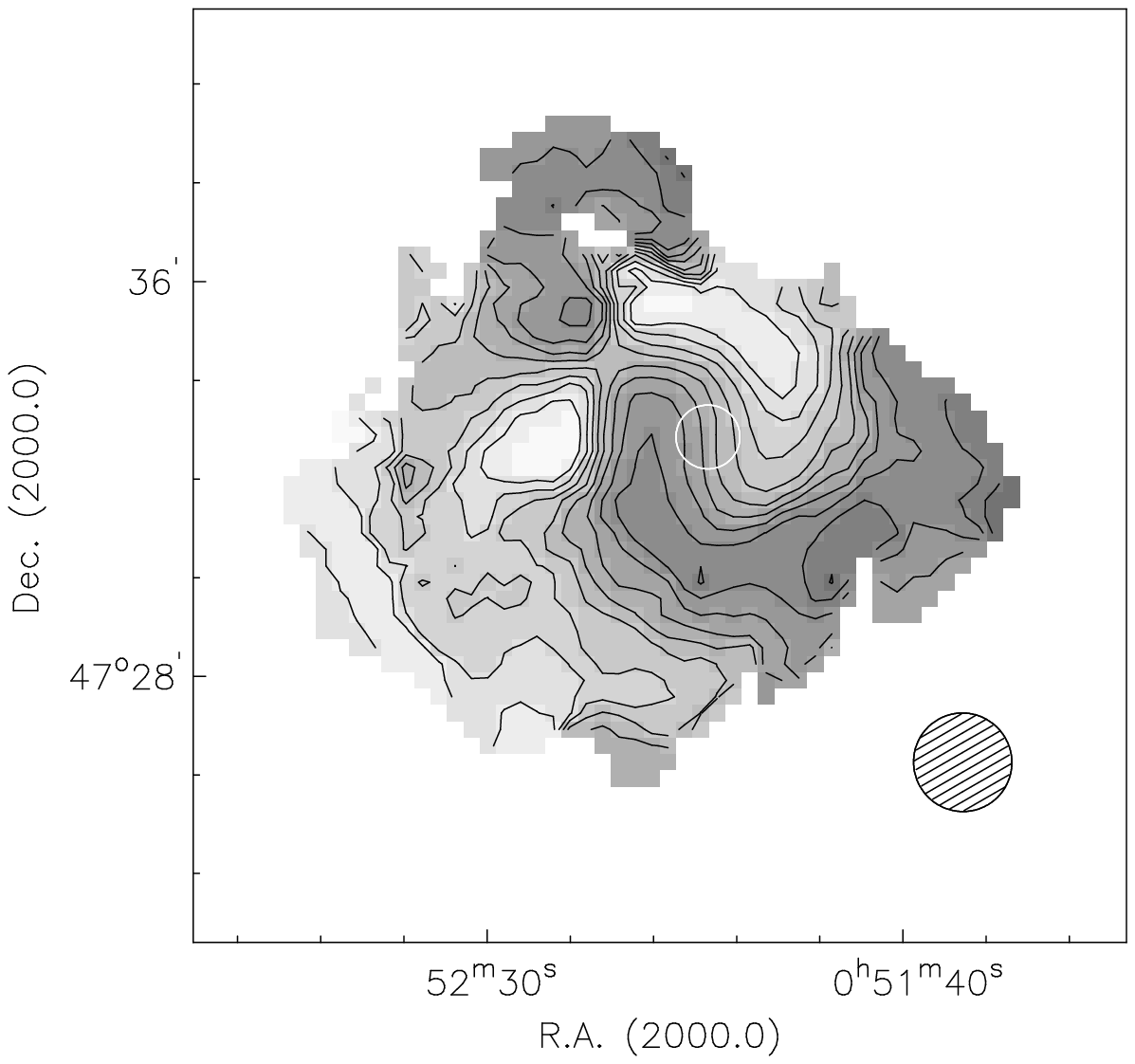,width=8cm,angle=-90}
\caption{Atomic hydrogen velocity field of NGC~278 at 120~arcsec
resolution. Contour levels are from 620\kms\ (light-grey) to 670\kms\
(dark-grey) in steps of 5\kms.}
\label{himom1}
\end{figure}

NGC~278 was observed as part of the WHISP project (van der Hulst, van
Albada \& Sancisi 2001; van der Hulst 2002).  One of the standard
products of the survey are data cubes at the full resolution of the
Westerbork Synthesis Radio Telescope (WSRT) array ($\sim15$~arcsec),
at 30~arcsec, and at 60~arcsec. In addition, we created a
low-resolution data cube with a beam size of 120~arcsec.  We produced
total surface density maps and velocity fields at each of the
resolutions.  We first isolated all signal over 2$\sigma$ in the
120~arcsec data, and, after removing spurious noise peaks, used this
as a mask for the 60~arcsec data in order to create the 60~arcsec
moment maps.  We then isolated all signal brighter than 2$\sigma$ in
the 60~arcsec masked map and used this as a mask for the 30~arcsec
data, and similarly for the full resolution data. We show the \hi\
surface density maps at 60 and 120~arcsec resolution in
Fig.~\ref{himom0}, and the velocity field at 120~arcsec resolution in
Fig.~\ref{himom1}.

The data covers the entire primary beam of the WSRT array ($\sim
30$~arcmin), and spans a velocity range from 380~\kms\ to 902~\kms\
with a channel separation of 4.14~\kms.  The rms noise in the
60~arcsec and 120~arcsec data which we use in this paper is
5~mJy\,beam$^{-1}$ and 7.6~mJy\,beam$^{-1}$, respectively.

\section{Results}

\subsection{Morphology}

\subsubsection{Optical morphology of the circumnuclear region and inner disk}

\begin{figure}
\psfig{figure=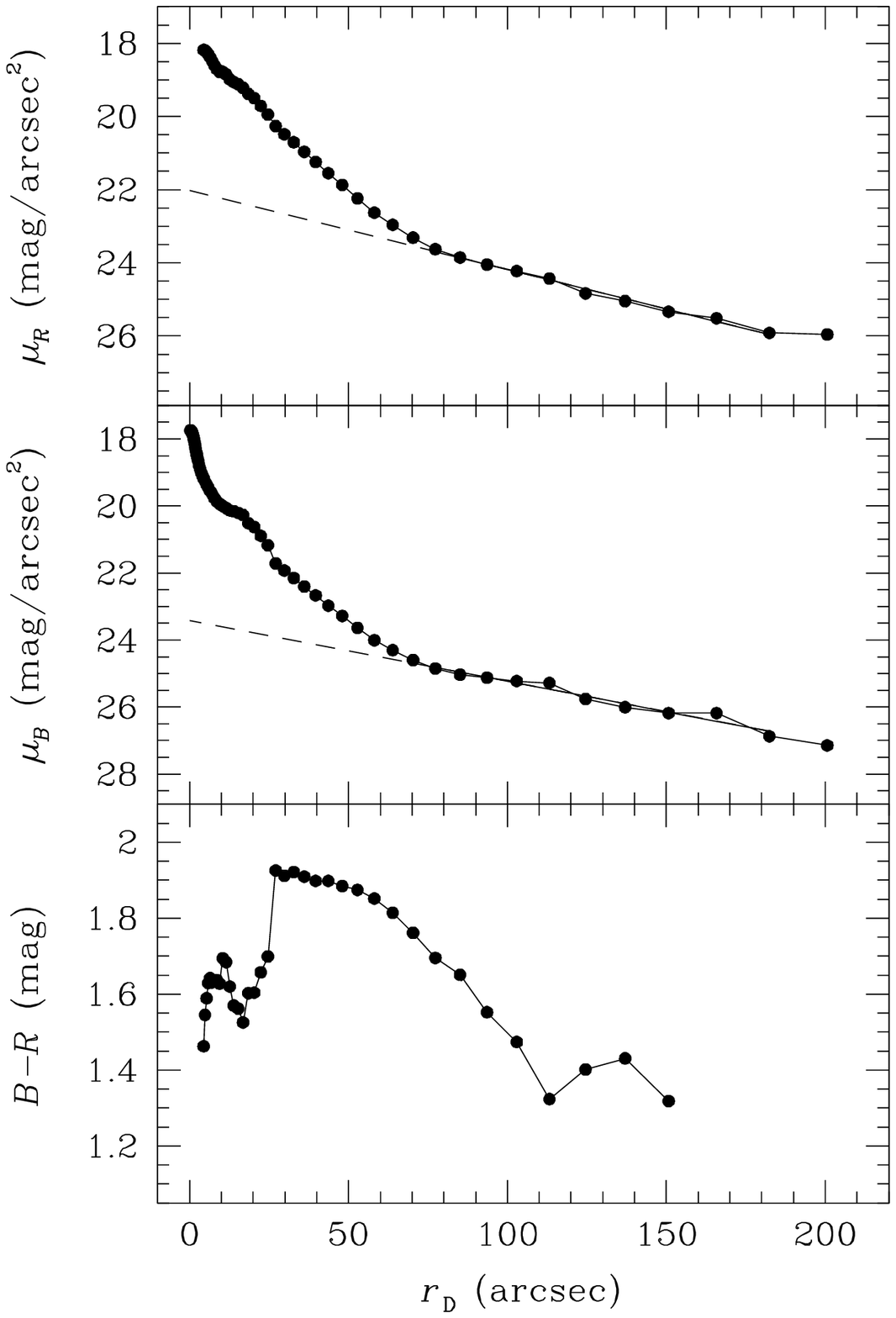,width=8cm}
\caption{Radial surface brightness profiles in the $R$ (top panel) and
$B$ (middle panel) bands, and radial $B-R$ colour index profile (lower
panel).  Least-squares fits to  the exponential  scale lengths  of the
outer  disk  in  $R$  and  $B$  are  indicated  by  the  solid  lines,
$h_R=43$~arcsec, $h_B=51$~arcsec, while the extrapolations of the fit inside
the fitted range in radius are shown by dashed line.}
\label{optprof}
\end{figure}

\begin{figure}
\psfig{figure=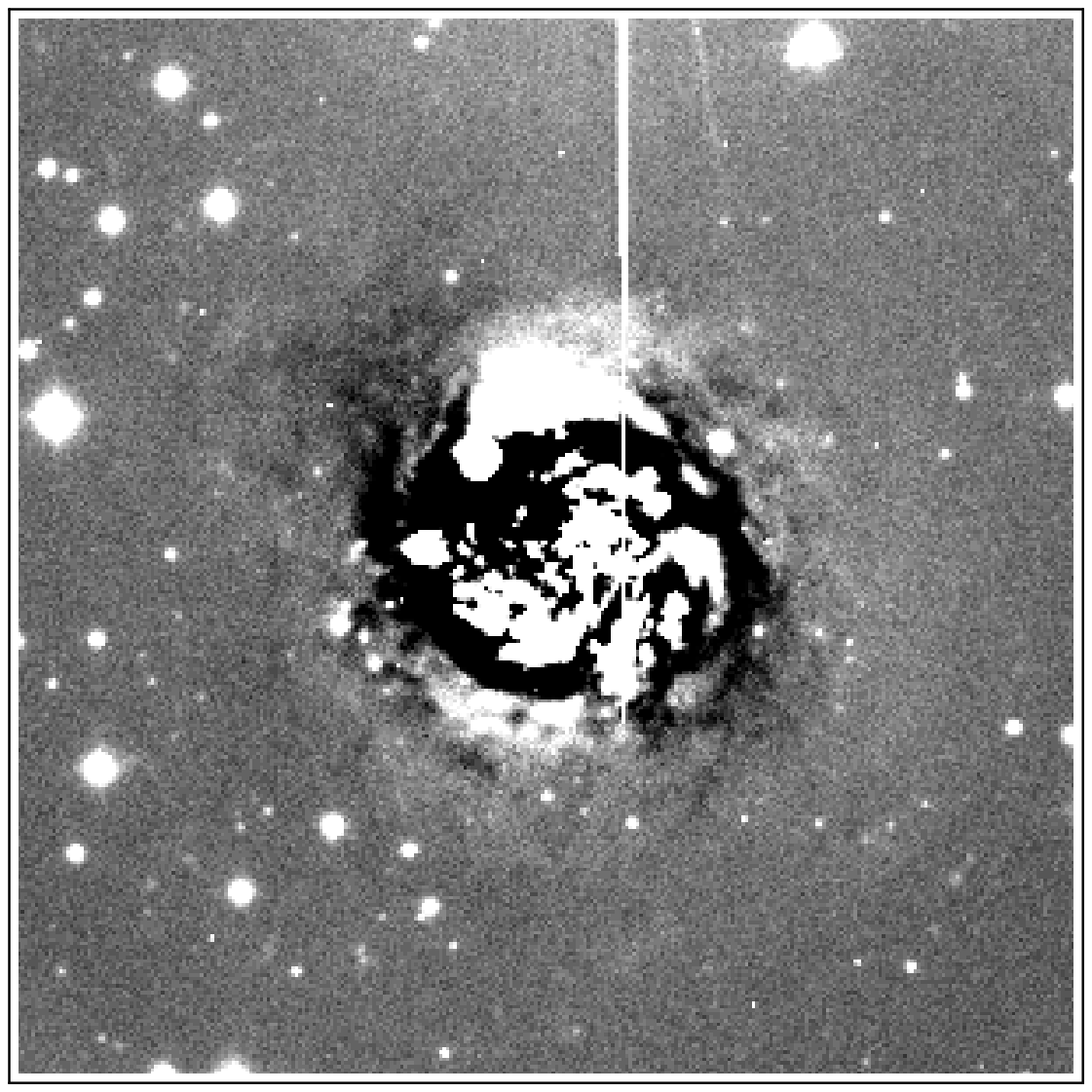,width=8cm}
\caption{$B$-band image of NGC~278 after subtracting an axisymmetric
model derived from the radial surface brightness profile
(Fig.~\ref{optprof}). Area shown is 3 arcmin on the side.}
\label{Bmodel}
\end{figure}

NGC~278 is characterised  by the strong dualistic behaviour  of the SF
in its  disk, with strong SF  in the inner 2~kpc  diameter region, and
hardly any SF at all outside  this sharply defined radius. This can be
seen in the images (Fig.~\ref{nir}, \ref{optical}, \ref{hst}) and most
clearly  in  the  radial  surface  brightness  and  especially  colour
profiles (Fig.~\ref{optprof}).  The latter  shows that within a radial
range of  some 5  arcsec, the  colour becomes redder  by at  least 0.3
mag. The bluer  colour within $r=26$~arcsec is caused  by enhanced SF, as
evidenced by, e.g., the \ha\ morphology (Fig.~\ref{moms}).

The outer disk, at radii larger than 27~arcsec or 1.5~kpc, is almost
completely featureless as best seen in optical broad bands
(Fig.~\ref{optical}).  A fit to the exponential disk in the radial
range from 100 to 180 arcsec gives a scale length of 51~arcsec in $B$
and 43~arcsec in $R$.  These fits are indicated as straight lines in
the radial surface brightness profiles (Fig.~\ref{optprof}).  Our \ha\
FP imaging shows a complete lack of massive SF activity outside the
sharp dividing radius of 27~arcsec. The spiral structure, so prominent
in the inner region, does continue into the outer region, but is very
weak and only shows up in the $B-R$ colour index image
(Fig.~\ref{optical}, lower left panel) and in the model-subtracted
$B$-band image of Fig.~\ref{Bmodel}.  To produce the latter, we
subtracted an azimuthally constant model, based on the radial surface
brightness profile, from the $B$ image. Both this and the colour index
image show how rather chaotic spiral structure can be followed
radially for at most another 1~kpc into the outer disk. This spiral
structure shows up in absorption: in red colour in the $B-R$ image,
and as dark or less emitting regions in the model-subtracted $B$
image.

\begin{figure}
\caption{$F450W - F814W$ {\it HST} colour index image. Darker shades
indicate redder colour.}
\label{hstcolour}
\end{figure}

\begin{figure*}
\caption{(a) (upper left) \ha\ FP velocity field of NGC~278 at 3
arcsec resolution. Grey-scale and contour levels are from 600 to
690\kms\ in steps of 10\kms, with the lower velocity contours in
black, and the first white contour at 650\kms. (b) (upper right) Model
velocity field as determined from the rotation curve (see
text). Contour and grey levels as in Fig.~\ref{modelvelfi}a. (c)
(middle left) Residual velocity map, obtained by subtracting the model
(Fig.~\ref{modelvelfi}b) from the velocity field
(Fig.~\ref{modelvelfi}a). Contours are at $-5, 0$ and 5\,\kms\ (black)
and $10, 15$ and 20\,\kms\ (white), with grey-scales indicating the
same range and higher values coded darker. (d) (middle right) As
Fig.~\ref{modelvelfi}c, now overlaid on a grey-scale representation of
the \ha\ total intensity, or moment zero, map at 3 arcsec resolution.
(e) (lower left) Position-velocity diagram along the kinematic major
axis of the central region (PA=65~degrees).  Contour levels are at
$-4$ and $-2\sigma$ (dashed), $2, 4, 8, 16$ and 32$\sigma$ (black),
and $64, 128$ and $256\sigma$ (white). Overlaid as black dots is the
rotation curve for the entire disk at the same resolution. The systemic
velocity of the galaxy is indicated by the horizontal dashed line. (f)
(lower right) As Fig.~\ref{modelvelfi}e, but now along the minor
axis.}
\label{modelvelfi}
\end{figure*}

The spiral structure in the inner disk, or within 1.5~kpc (27~arcsec)
in radius, is prominent but rather flocculent (Fig.~\ref{hst},
\ref{hstcolour}). The nuclear ring of radius $\sim$5~arcsec identified
by Pogge (1989) can be seen as an \ha\ ring in the FP moment zero maps
(Fig.~\ref{moms}, \ref{modelvelfi}d).  It seems to be slightly offset
from the kinematic centre of the galaxy.  This nuclear ring can also
be recognised on the {\it HST} broad-band or colour index images, as
regions with enhanced SF.  Both the \ha\ total intensity map
(Fig.~\ref{moms}) and the colour index images (Fig.~\ref{optical})
show that the SF activity within the inner disk area is biased toward
the outer regions.  This, coupled with the whispy spiral structure
outlined in dust (red colours) in the colour index images, most
spectacularly so in the {\it HST} image, leads to a similarity of
NGC~278's inner region to nuclear ring regions in galaxies like M100
or NGC~5248 (e.g., Knapen et al. 1995a,b; Jogee et al.  2002a,b; see
Buta \& Combes 1996 for a review on rings).  We will discuss this
nuclear ring in more detail in Sect.~4.5.

\subsubsection{\hi\ morphology of the outer regions}

The atomic hydrogen distribution  of NGC~278 can be distinguished into
two components:  a compact (though  in itself larger than  the optical
disk of the galaxy) central  region of high column density, surrounded
by an extended complex of \hi\ clouds at low column density. The central
part is readily visible at all resolutions, but the outer component is
most clearly seen in the 120~arcsec and 60~arcsec resolution maps.

Figure~\ref{himom0} shows the integrated surface density maps of
NGC~278 at 60~arcsec and 120~arcsec resolution.  The column densities in
the outer and inner parts differ by over a factor of 10, with a sharp
increase at the boundary between the two components. The extent of the
optical component of NGC~278, as visible on a DSS image, is indicated
in Fig.~\ref{himom0} as a white circle.  The comparison illustrates
clearly that the \hi\ disk is very extended compared to the optical
distribution.  Due to the low spatial resolution needed to see the
outer \hi\ emission, it is hard to determine the detailed structure of
the atomic hydrogen, but it seems from Fig.~\ref{himom0} that the \hi\
is organised somewhat, possibly in arm fragments, of spiral or tidal
origin.

\begin{figure}
\psfig{figure=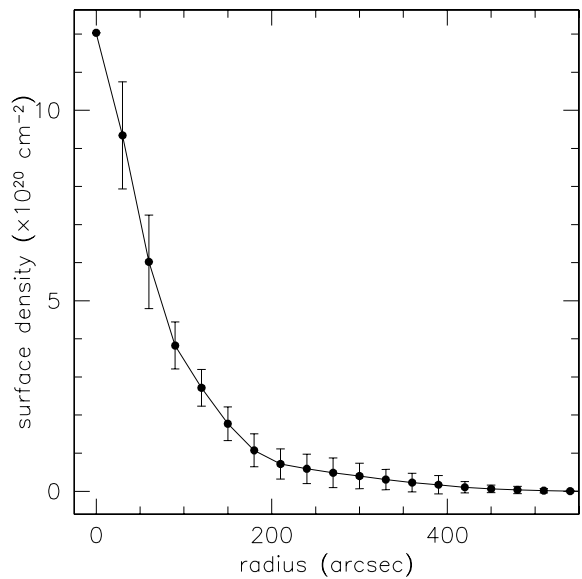,width=8cm}
\caption{\hi\   surface  density   profile  of   NGC~278   at  60~arcsec
resolution.}
\label{hiprof}
\end{figure}

These results  are illustrated  by Fig.~\ref{hiprof}, which  shows the
\hi\ surface density profile of  the 60~arcsec data, derived assuming an
inclination of 17~degrees and  a position angle (PA) of $-50$~degrees,
which is the  approximate PA of the kinematic major  axis (at such low
inclinations the  precise value is not critical).   The position where
the \hi\ distribution peaks was adopted as the centre.

The total flux of NGC~278 as derived from the 120~arcsec data is
31.7\,Jy\,\kms, which agrees well with the single-dish value of
29.8\,Jy\,\kms, as given by Huchtmeier \& Richter (1989). The
corresponding \hi\ mass is $1.1\times10^9$\,M$_\odot$. The mass of
the bright central part (i.e., inside 180~arcsec) is
$7.3\times10^8$\,M$_\odot$, that of the outer component
$3.3\times10^8$\,M$_\odot$.  The outer components thus contains some
30\% of the total atomic hydrogen mass of the system.

\subsection{Kinematics}

\subsubsection{\ha\ kinematics in the inner region}

The velocity field of the inner disk region in NGC~278 (the only
region available using \ha\ as a tracer, given the absence of \ha\
emission outside $r=27$~arcsec) is dominated by ordered circular
motion (Fig.~\ref{moms}, \ref{modelvelfi}).  Prominent non-circular,
or rather non-regular, motions are seen, however, both in the form of
the general clockwise rotation of the isovelocity contours toward the
outer end of the area surveyed in \ha, and in the form of small-scale
deviations well within that area.  The spatial coincidence of the
small-scale deviations from circular motion with the spiral arms as
traced by the integrated \ha\ emission (Fig.~\ref{moms},
\ref{modelvelfi}) suggests that these motions are induced by density
wave streaming across the spiral arms.  The PA of the kinematic minor
axis in the central part of NGC~278 can be estimated from the velocity
fields as 155~degrees. Closed isovelocity contours indicate that the
circular rotation velocities reach a local maximum near a radius of
18~arcsec.

\begin{figure}
\psfig{figure=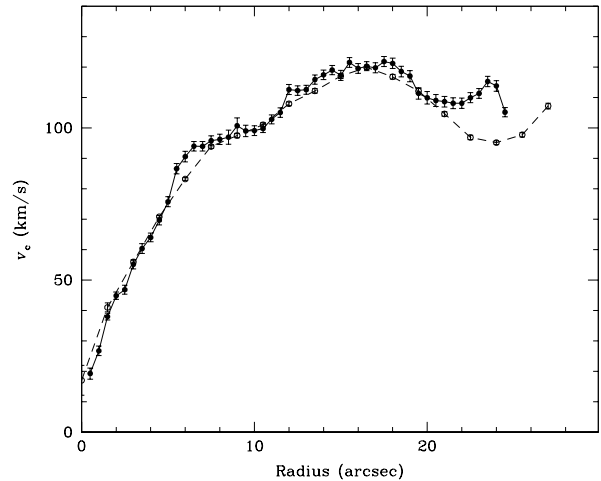,width=8cm}
\caption{Rotation curve as derived from the  1.2~arcsec (filled symbols
and drawn line) and from the 3~arcsec (open symbols and dashed line) spatial
resolution H$\alpha$ velocity field of NGC~278.}
\label{rc}
\end{figure}

We derived rotation curves from the 1.2 and 3 arcsec resolution
velocity fields, following the description of Knapen (1997). Because
the galaxy is nearly face-on, the inclination angle cannot reasonably
be constrained by the fitting procedure, and we assumed a constant
inclination of 17~degrees.  The resulting rotation curves are shown in
Fig.~\ref{rc} and show a relatively slow rise, and the local maximum
of $\sim120$\kms\ (or more accurately, 35/$\sin i$\kms) at
$r=18$~arcsec already seen from the velocity fields. The rotation
curve at 3~arcsec resolution is also shown overplotted on the major
axis position-velocity plot, in Fig.~\ref{modelvelfi}e.

We used the latter curve, along with the fitted radial run of major
axis PA (not shown), to produce a model velocity field, which can be
assumed to trace circular motions. This model velocity field is shown,
along with the original 3~arcsec resolution velocity field from which
it was derived, in Fig.~\ref{modelvelfi} (panels b and a,
respectively).  Subtracting this model from the observed velocity
field highlights the location and approximate amplitude of the
residual, or non-circular, motions.  This is shown in
Fig.~\ref{modelvelfi}c, and overlaid on the total intensity \ha\ map
in Fig.~\ref{modelvelfi}d.  As described further by Knapen et
al. (2000), the fact that the rotation curve used for producing the
model velocity field contains a certain contribution from the
non-circular motions can lead to an underestimate of the non-circular
motions in this technique.  Here we will limit ourselves, however, to
a qualitative discussion and this can be well based on the residual
velocity field, as shown in Fig.~\ref{modelvelfi}.

This analysis shows that the clockwise skewing of the isovelocity
contours near the outer end of the region observed in \ha\ can be
reproduced with circular motions, albeit significantly changing the PA
of the line of nodes.  Because the effect occurs where the
signal-to-noise ratio in the \ha\ data is lowest, it must be treated
with care, but in comparison to the non-regular \hi\ velocity field
(Sect.~3.2.2) it is entirely reasonable to suspect that its origins
lie in the same process that causes the \hi\ velocity field to be so
irregular, presumably a recent interaction as we will see
below. Another possible explanation for the shape of the \ha\ velocity
field could be the presence of a bar, but given that our imaging shows
no evidence whatsoever of such a presence, this interpretation is
unlikely.

Concentrating now on the regions of non-circular motion within the
area mapped in \ha, the residual velocity field shows, in white
contours, that the most prominent of these regions occur to the SE, to
the W and the ENE of the nucleus.  Comparison with the \ha, optical
and NIR images, but especially with the {\it HST} image
(Fig.~\ref{hst}) shows that this is where the most prominent spiral
arm patches are located.  This hints at an origin of the non-circular
motion in density-wave induced streaming of gas across the spiral
arms. The amplitude of the non-circular motions, of some $10-15/\sin
i$\,\kms, or $34-51$\,\kms, is fully consistent with this
interpretation, and although the spiral pattern may not at first sight
look grand-design, colour index maps (Fig.~\ref{optical},
\ref{hstcolour}), as well as the unsharp-masked image of
Fig.~\ref{Bmodel}, do show evidence for bi-symmetric spiral structure.

\subsubsection{\hi\ kinematics in the outer disk}

We  investigated  the  atomic   hydrogen  kinematics  at  all  available
resolutions, but  found that  the signal to  noise ratio in  the outer
parts is too  low to construct a meaningful velocity  field in all but
the lowest-resolution data sets. The  \hi\ velocity field of NGC~278 at
120~arcsec  resolution is  shown  in Fig.~\ref{himom1}.  We stress  that
comparison with the high resolution  (1~arcsec) velocity field in \ha\ is
nearly  impossible: the  area mapped  in \ha,  of at  most  30~arcsec in
radius, is in its entirety well within the area covered by the central
beam in the \hi\ data.

In the \hi\ velocity field, as in the \hi\ column density maps, we can
also distinguish two components: the inner part, corresponding to the
high column density central region, shows a well-ordered velocity
field, indicating a rotating disk, though with possibly a strong warp.
Outside a radius of $\sim3$~arcmin, the velocity field changes
dramatically, and hardly any ordered rotation can be
distinguished. The \hi\ in these outer parts, however, does seem to
form a coherent structure.  Especially in the Eastern region, we see a
continuity in velocity between the Northern and Southern parts, which
suggests that this part of the \hi\ distribution constitutes in
NGC~278 something like a tidal stream.  Unfortunately, the higher
resolution data do not have enough S/N to verify this idea in detail.
There is also an intriguing continuity in velocity between the western
part of the inner disk and the outer disk, perhaps suggestive of an
infalling cloud.  Given the absence of ordered motions, it is not
meaningful to try and deduce a rotation curve, and by implication no
information on the halo of this galaxy can be extracted from the
velocity field. Some aspects of the \hi\ velocity field are
reminiscent of the effects a bar can have on an atomic hydrogen
velocity field (e.g., Bosma 1980). As we will see below, however, it
is much more likely that the deviations from circular motion, as well
as the \hi\ morphology and in fact the nuclear ring of SF, are a
result of a past interaction.

Below, we will describe how the \hi\ morphology and kinematics fit
into our proposed scenario of a recent interaction as the origin of
the currently observed structure of NGC~278.

\section{Discussion}

\subsection{Summary of observational results}

Before we discuss a plausible history of NGC~278, we briefly review
the main observational findings. Together, these results outline the
puzzle presented to us by this small galaxy.

\begin{enumerate}

\item NGC~278 possesses a large \hi\ disk, with possibly in the outer
regions the remains of a previously caught, gas-rich, companion.

\item In the central region, an area of active massive SF is found.

\item Rather red colours in the optical disk may indicate an advanced
age, and/or significant extinction.

\item The kinematics, both at small and large scales, are severely
disturbed. In the outer disk, well outside the optical disk, the
disturbance may be due to large-scale deviations from axisymmetry.

\item The disk has a low surface brightness, but its colour is much
redder than the typical disk found in late-type low surface brightness
galaxies (LSB galaxies).

\item The circumnuclear \ha\ morphology outlines a nuclear disk, which
has a rather typical size in absolute terms, but is unusually large in
comparison to the disk scale length or the size of its host galaxy.

\end{enumerate}

In the following sections we will discuss the main features of this
galaxy in view of the properties mentioned above. We will consider the
surface brightness of the disk, the bright central part which shows
little evidence for a bar, and the features directly related to the
possibility of NGC~278 being the remnant of an accretion event: the
large \hi\ disk and the bright nuclear ring of SF.

\subsection{The low surface brightness disk of N278}

The disk of NGC~278 has a very low surface brightness: the
extrapolated central value in the $B$-band
($\mu_{0,B}=23.5\,$mag\,arcsec$^{-2}$) is comparable to that typically
found for late-type LSB galaxies ($\mu_{0,B} \simeq
23.2\,$mag\,arcsec$^{-2}$; de Blok, van der Hulst \& Bothun 1995). As
it has such a low surface brightness disk, a possible explanation for
the current evolutionary state of NGC~278 could be that it is an LSB
galaxy with much enhanced SF in the central few kiloparsec region,
organised into spiral arms.

This is, however, not a likely scenario as the $B-R$ colour of the
outer part of the optical disk of NGC~278 is about 1.5~mag (see
Fig.~\ref{optprof}), redder than the average LSB galaxy which has a
$B-R$ colour of 0.78 (de Blok et al. 1995). This is a significant
difference given the small spread in $B-R$ exhibited by the LSB
galaxies.

Comparing the LSB galaxies from de Blok et al. (1995) and the HSB
galaxies from de Jong \& van der Kruit (1994) we find that NGC~278 is
in a region of the $B-R$ vs.  surface brightness diagram (e.g., fig.~8
in de Blok et al. 1995 or fig.~2.7 in de Blok 1995) which is only
populated by a few faint galaxies from the de Jong \& van der Kruit
sample.  So the disk of NGC~278, despite its low surface brightness,
must have a different SF history than the late-type LSB galaxies
studied by de Blok et al. (1995).

\subsection{No bar}

The purpose of this short Section is to reiterate that there is no
evidence for the presence of a bar in NGC~278. Although the galaxy has
been classified as ``SAB'' in the RC3, we see no evidence from our NIR
(Fig.~\ref{nir}) or optical (Fig.~\ref{optical}) imaging. The NIR
images only cover the innermost region of the disk, up to a radius of
some 30~arcsec, and although some deviations from perfectly circular
isophote shapes can be seen in the Figure, these can, upon further
inspection, be traced to localised regions of SF. We fitted ellipses
to the NIR isophotes (not shown) and confirm from these what
Fig.~\ref{nir} shows qualitatively: there is no evidence for the
presence of a bar in the area covered by the NIR images.

The optical images, shown in Fig.~\ref{optical}, cover a somewhat
larger area of the disk, but also show absolutely no evidence for a
bar from the isophotes (Fig.~\ref{optical}), nor indeed from ellipse
fitting (not shown).

There are two other possibilities, both of which can be discarded as
being rather implausible.  The first is that a large bar is present,
completely outside the range of our optical imaging, but which affects
the velocity field of the outer part of the disk as seen in \hi. Since
such a bar would lie completely below a surface brightness level of
26\,mag\,arcsec$^{-2}$ in this nearby galaxy, it is not an
attractive scenario, although we cannot formally rule it out.  The
second possibility is that there is a tiny nuclear bar which we cannot
detect with our current imaging. {\it HST} NIR imaging is not
available for NGC~278, but the WFPC2 optical imaging (Fig.~\ref{hst})
shows no indication whatsoever for the presence of a mini-bar. We
conclude that there is no evidence for the presence of a bar in
NGC~278.

\subsection{Possible merger history of NGC~278}

Following the prescription of Laine et al. (2002), we checked in the
Lyon-Meudon extragalactic database (LEDA) whether NGC~278 has any
nearby companions. Using a search radius of 400~kpc around the galaxy,
and imposing a $\pm500$ km~s$^{-1}$ range in $cz$, we found no
companions. Extending our search to a wider area around NGC~278, we
found the galaxy UGC~672, at $v_{\rm sys}=708$\,\kms, already
discussed in the introduction (Sect.~1). This galaxy, however, is
faint, with a magnitude of 18 given in NED, which would correspond to
an absolute magnitude $M\sim-12.4$\,mag.  It is also at a distance of
some 600~kpc from NGC~278, and can thus not be considered a companion
that could have had any effect on the evolution of NGC~278. NGC~278 is
thus isolated, at least at present.

We have searched the \hi\ data cubes for possible companions but did
not detect any. This puts rather strong limits on the presence and
masses of any possible companions, as well as on interaction
timescales.  Given that the noise in the 120~arcsec data is
7.45~mJy\,beam$^{-1}$, and assuming that any undetected galaxies in
the field have a 3$\sigma$ peak flux upper limit and a profile width
of 50~\kms, the flux of such an undetected galaxy would be
1.11~Jy\,\kms, which at the distance of NGC~278 corresponds to an
upper limit in \hi\ mass of 8.9$\times10^5$\,M$_\odot$. The 3$\sigma$
column density limits in the 120~arcsec and 60~arcsec data are
7.1$\times10^{18}$\,cm$^{-2}$ and 2.1$\times10^{19}$\,cm$^{-2}$,
respectively.

The \hi\ morphology and kinematics, however, especially in the outer
parts of the disk, do yield evidence for a recent interaction. Both
the asymmetrical morphology and disturbed kinematic structure of the
disk are reminiscent of tidal tails and arms as found in numerical
simulations of galaxy interactions (dating back to the classic paper
by Toomre \& Toomre 1972).

Thus even though NGC~278 appears isolated at present, there is
convincing evidence for a recent interaction. We speculate that this
interaction may have been with a gas-rich dwarf, which has ceased to
exist as such, and whose gaseous material now forms part of the tidal
arm in the outer disk of NGC~278. The interaction may well have been
of a similar nature to those studied by Hernquist \& Mihos (1995), who
considered the case of a minor merger between a gas-rich disk and a
dwarf galaxy, and who found that such a merger can lead to torquing
and subsequent radial inflow of gas in the primary disk.  The total
atomic gas mass of NGC~278 at present, of $1.1\times10^9$\,M$_\odot$,
would then be the total of the masses of the two individual galaxies
before the collision, which implies that both must have been rather,
but by no means implausibly, small galaxies.  Assuming, as above, an
absolute magnitude $M_B = -18.8$ implies that NGC~278 has an \hi\ mass
to luminosity ratio $M_{\rm HI}/L_B \sim 0.4$. Such a ratio is about
twice as high as expected for a typical Sb galaxy (Roberts \& Haynes
1994), and more typical of late-type Sd galaxies. This is an
indication that significant gas infall has taken place at some time in
the past. In view of the red colour of the disk, which indicates an
advanced age, the event under consideration here has perhaps been the
first merger event undergone by this galaxy.

\subsection{A nuclear ring in a non-barred galaxy}

Perhaps the most intriguing feature of this small spiral galaxy,
NGC~278, is the contrast between the 2.2~kpc diameter central region,
with strong massive SF activity and well defined spiral structure, and
the region outside 27 arcsec in radius, where a mostly featureless
disk without any noticeable SF extends to about 8~kpc in radius in the
optical, and much further as seen in \hi. The break between these two
regions is caused by the sudden outward demise of SF and is abrupt, as
seen in the images (Figs.~\ref{nir}, \ref{optical}, \ref{hst}) and in
the radial profiles, especially of colour (Fig.~\ref{optprof}).  The
spiral structure continues a bit into the outer disk, but in
absorption (presumably caused by dust) more than in emission
(Fig.~\ref{hstcolour}, \ref{Bmodel}).  The \ha\ image
(Fig.~\ref{moms}) and the radial colour profile, which suddenly
becomes significantly redder (Fig.~\ref{optprof}), constitute proof
that the change in appearance between the inner and outer regions is
indeed due to SF alone.

It is tempting to consider a star formation threshold defined by a
Toomre criterion for the onset of star formation, following the
analysis of Kennicutt (1989; 1998). But firstly, the drop in
intensity, not just in the massive star formation tracer \ha\ but also
in underlying optical light, is very large, larger than normally seen
in spiral galaxies. Secondly, a Kennicutt-style analysis requires the
observation of a rotation curve in order to calculate a critical gas
density, and of a gas density profile to compare to the star formation
profile. In NGC~278, the velocity field is so different from
well-ordered circular rotation that a rotation cannot be derived,
whereas the edge of the star formation is well within the central beam
of our neutral gas data.

As noted above, in Sect.~3.1, the appearance of the inner region,
especially in colour index images and in the {\it HST} real-colour
image (Figs.~\ref{nir}, \ref{optical}, \ref{hst}) is reminiscent of
star-forming nuclear rings in barred galaxies. Such nuclear rings are
generally interpreted as dynamical effects of gas flow in the presence
of ILRs which halt the net radial inflow of gas under the influence of
a bar (e.g., Knapen et al. 1995a).

There are, however, a number of interesting differences between the
nuclear ring in NGC~278 and those in other galaxies.  First and
foremost, the presence of nuclear rings is in most cases directly
related to the presence of a bar (e.g.  Buta \& Combes 1996), yet
NGC~278 shows no trace of a bar. Second, the inner region in NGC~278
is relatively large for a nuclear ring. This is not so much the case
for the diameter of the ring itself (2.2~kpc), but rather in
comparison to the size of the disk. The ring diameter is comparable
to, or slightly larger than, the scale length of the exponential disk
of the galaxy which hosts it, which is in stark contrast with the
typical behaviour of nuclear rings, where the scale length is $3-10$
times larger than the ring diameter (e.g., Knapen, P\'erez-Ram\'\i
rez \& Laine 2002).  So whereas for NGC~278 $h/D_{\rm ring}\sim1$,
for instance, for NGC~4314 this ratio is 6.4, for NGC~4321, 3.4, and
for NGC~5248, 3.7 (Knapen et al. 2002). Another measure of ring size
is its size relative to the size of the complete disk ($D_{25}$). For
NGC~278, $D_{\rm ring}/D_{25}$ is 0.31, twice as large as the largest
nuclear rings among the sample of 15 compiled by Knapen (2004).  We
thus find that the nuclear ring in NGC~278 is a relatively large
nuclear ring, hosted by a non-barred, and extremely small disk galaxy.

Buta \& Combes (1996) review the existence of rings (not only nuclear
ones) in non-barred galaxies, and state that for most cases their host
galaxies present evidence for some other kind of mechanism which sets
up a non-axisymmetric potential, for instance a weak oval distortion,
or the tidal action of a bound companion. We believe that this is also
the case here, where our favoured explanation for the origin of the
nuclear ring in NGC~278 is a non-axisymmetric potential, set up as a
result of the minor merger with a companion. This potential then has a
very similar effect to one set up by a classical bar, and can result
in a nuclear ring.  NGC~278 thus represents a case for the effects of
interactions on the gas distribution within galaxies, and for the
formation of a nuclear ring in a {\it non-barred} galaxy.

Finally, NGC~278 hosts a small ring inside the inner region, as seen
in \ha\ (Sect.~3.1), at a radius of 0.2~kpc.  We have found a very
similar case in the galaxy NGC~5248.  There, a large bar hosts a set
of two rings as seen in \ha\ (Laine et al. 2001; Maoz et al. 2001), at
radii of 0.45 and 0.1~kpc (Jogee et al. 2002b) (where the scale length
of the exponential disk is 3.3~kpc; Knapen et al.  2002).  Such
mini-rings may be more common than is now recognised because they can
only be found from high resolution imaging of nearby galaxies,
preferably in emission lines which trace massive SF.  Possible origins
for such mini-rings can lie in the dynamical structure set up by the
host galaxy, similar to classic nuclear rings forming between a pair
of ILRs, or in SF triggered by outflow resulting from nuclear
starburst or even non-stellar activity.

\section{Conclusions}

Using a combination of FP \ha\ data, optical and NIR imaging, and
atomic hydrogen mapping, we make a detailed study of the small spiral
galaxy NGC~278.  The main conclusions from our work can be summarised
as follows.

\begin{enumerate}

\item NGC~278 is a small face-on spiral galaxy, of optical diameter
only 7.2 kpc. Most of this disk is not actively forming stars, and the
SF is concentrated within the central 1.5~kpc radius.

\item The massive circumnuclear SF is organised in a nuclear ring,
which contains a set of patchy spiral arms.  The \ha\ velocity field
shows some streaming motions due to these spiral arms.

\item The nuclear ring in NGC~278 is not particularly large in
absolute terms (radius of one kpc) but is at least twice as large as
other nuclear rings when its size relative to that of the host galaxy
is considered, or its size relative to the scale length of the disk of
its host galaxy.

\item We find no evidence for the presence of a bar in NGC~278, at any
scale.

\item At present, NGC~278 does not have any companions which can
gravitationally affect it.

\item  The \hi\  velocity field shows  large deviations  from circular
motion. We also observe an  irregular structure of the atomic hydrogen
morphology,  and interpret  these findings  as evidence  for  a recent
merger with a  small gas-rich galaxy.

\item We postulate that this interactive event has led to a
non-axisymmetry in the gravitational potential, which in turn has
facilitated the formation of the nuclear ring. This explains why a
nuclear ring can exist in a non-barred galaxy.

\end{enumerate}

\begin{acknowledgements}

We thank Seppo Laine for his help with the reduction of the FP data,
Jim Lewis for help with the {\sc taucal} reduction software, and
Dolores P\'erez-Ram\'\i rez, Ren\'e Doyon and Daniel Nadeau for their
help with the observations. Sharon Stedman kindly fitted the radial
profiles to the INT data, and Nik Szymanek produced the real-colour
image of NGC~278 (Fig.~\ref{hst}). We thank Isaac Shlosman and Elias
Brinks for comments on an earlier version of this manuscript. LFW
thanks the Department of Physical Sciences of the University of
Hertfordshire (UH) and the ING for financial help for a visit to the
ING where this work was advanced.  The present paper includes results
from a final year project by LFW at UH, for which support from UH is
acknowledged.  The William Herschel and Isaac Newton Telescopes are
operated on the island of La Palma by the ING in the Spanish
Observatorio del Roque de los Muchachos of the Instituto de Astrof\'\i
sica de Canarias.  The WHISP project has been supported by the Kapteyn
Institute and the Foundation for Research in Astronomy (ASTRON) and
has been carried out with the WSRT. The WSRT is operated by ASTRON
with financial support from the Netherlands Organisation for
Scientific Research (NWO).  We used data obtained with the
Canada--France--Hawaii Telescope, operated by the National Research
Council of Canada, the Centre National de la Recherche Scientifique de
France and the University of Hawaii.  Partly based on observations
made with the NASA/ESA {\it Hubble Space Telescope}, obtained from the
data archive at the Space Telescope Science Institute, which is
operated by the Association of Universities for Research in Astronomy,
Inc., under NASA contract NAS 5-26555.

\end{acknowledgements}

\end{document}